\documentclass[twocolumn,prl,floatfix,aps,showpacs]{revtex4-1}
\usepackage[utf8]{inputenc}
\usepackage{amsmath}
\usepackage{graphicx}

\makeatletter

\usepackage{color}

\pdfpageheight\paperheight
\pdfpagewidth\paperwidth

\usepackage{subfigure}
\usepackage{dcolumn}
\usepackage{bm}
\usepackage[bb=boondox]{mathalfa}
\usepackage{comment}

\usepackage{siunitx}
\usepackage{physics}
\usepackage{pgfplots}
\usepackage{tikz}\usepackage{tcolorbox}

\usepgfplotslibrary{fillbetween}
\usetikzlibrary{arrows,patterns}
\pgfplotsset{compat=1.10}

\newcommand{\phix}[1][]{\phi_{#1}(x,t)}

\newcommand{\e}[1]{e^{#1}}

\newcommand{\phitt}[2][]{\tilde{\phi}_{#1}(#2)}
\newcommand{\pphi}[2][]{\phi_{#1}(#2)}
\newcommand{\nn}{\nonumber}
\newcommand{\px}{\partial_x}
\newcommand{\pt}{\partial_t}
\newcommand{\gp}{\gamma_+}
\newcommand{\gm}{\gamma_-}

\DeclareMathOperator{\Ei}{Ei}

\newtcolorbox{mybox}[1][]{colback=white, sharp corners, #1}

\newcommand{\tauRC}{\tau_{\textrm{\tiny RC}}}

\makeatother

\begin{document}

\title{Tunneling into a finite Luttinger liquid coupled to noisy capacitive
leads}

\author{Antonio Štrkalj}

\author{Michael S. Ferguson}

\author{Tobias M. R. Wolf}

\author{Ivan Levkivskyi}

\author{Oded Zilberberg}

\affiliation{Institute for Theoretical Physics, ETH Zurich, 8093 Zurich, Switzerland}

\date{\today}
\begin{abstract}
Tunneling spectroscopy of one-dimensional interacting wires can be
profoundly sensitive to the boundary conditions of the wire. Here,
we analyze the tunneling spectroscopy of a wire coupled to capacitive
metallic leads. Strikingly, with increasing many-body interactions
in the wire, the impact of the boundary noise becomes more prominent.
This interplay allows for a smooth crossover from standard 1D tunneling
signatures into a regime where the tunneling is dominated by the fluctuations
at the leads. This regime is characterized by an elevated zero-bias
tunneling alongside a universal power-law decay at high energies.
Furthermore, local tunneling measurements in this regime show a unique
spatial-dependence that marks the formation of plasmonic standing
waves in the wire. Our result offers a tunable method by which to
control the boundary effects and measure the interaction strength
(Luttinger parameter) within the wire. 
\end{abstract}
\maketitle
Advances in control and design of mesoscopic systems have made it
possible to realize a variety of ultra-small electronic tunnel-junctions
\cite{nazarov1992,ihn2004}. In such junctions, many-body interactions
and coherent effects compete with the charge fluctuations and impedance
of the environment to profoundly impact the resulting tunneling characteristics;
the tunneling inside the junction excites the electromagnetic modes
of an external circuit making it extremely sensitive to the circuit's
impedance \cite{nazarov1992,ihn2004,devoret2005}. This competition
alters the tunneling density of states (TDOS) of the various device
constituents, with a wide variety of such effects seen in, e.g., normal-metal
tunnel-junctions~\cite{nagae1971}, Josephon junctions~\cite{josephson1962}
and transmission lines~\cite{chakravarty1986}. Particular examples
of such effects include, among others, the Coulomb blockade \cite{kouwenhoven2001},
the Kondo effect \cite{goldhaber1998,kouwenhoven_glazman2001,oehri2015}
and Andreev bound modes \cite{andreev1964,Woerkom2017aa,Nichele2017,Das2012aa}.

Tunnel-junctions involving one-dimensional (1D) quantum wires are
especially intriguing, since many-body interactions fundamentally
alter the emergent many-body physics compared with conventional Fermi-liquid
metals. Interacting wires are better described using Tomonaga-Luttinger
liquid (TLL) theory \cite{tomonaga1950,luttinger1963,haldane1981}:
the low-energy elementary excitations in 1D appear as collective bosonic
plasmon modes --- in stark contrast to the constitutive fermionic
electrons. Consequently, 1D systems show exotic phenomena, such as
charge fractionalization of injected electrons \cite{safi1995,ivan2016},
spin-charge separation~\cite{auslaender2005,jompol2009}, and zero-bias
anomalies (ZBA) \cite{kane_fisher1992,mateev_glazman1993,glazman2001,gutman2008},
all of which uniquely interplay with disorder~\cite{apel_rice1982,giamarchi_schulz1988},
quasi-disorder~\cite{giamarchi2000}, and dissipation~\cite{altland_gefen2012,altland_gefen2015}.
Such 1D effects are ubiquitous and have been observed in a wide variety
of systems, including nanotubes \cite{bockrath1999,yao1999}, GaAs
wires \cite{auslaender2005,jompol2009}, quantum Hall edges \cite{wen1990,Chang2003,heiblum2003},
as well as, chains of spins or atoms \cite{blumenstein2011,brantut2015}.

More recently, significant progress was made in the description of
realistic finite-sized 1D wires with boundary conditions both in-
and out-equilibrium \cite{giamarchi,ivan2012,gutman2008,gutman2009,gutman2010}.
These can generally be grouped into wires (i) with open boundaries
\cite{eggert1996,eggert2008,nazarov1997}, (ii) connected to ohmic
contacts \cite{ivan2013}, or (iii) coupled to inherently out-of-equilibrium
charge distributions \cite{gutman2008,gutman2009}. Interestingly,
despite the fact that the many-body interactions profoundly alter
the emergent quasiparticle excitations in the wire relative to the
electronic boundaries, the wire--boundary interplay cannot be revealed
in DC-transport measurements due to the suppression of electron backscattering
in clean wires \cite{kane_fisher1992_2,giamarchi,matveev2010}. In
contrast, a tunnel-junction between a superconducting or metallic
scanning tunneling microscope (STM) and the wire is ideally suited
to sense these effects, since it gives access to the wire's energy
distribution function \cite{pothier_devoret1997,anthore_pierre_pothier_esteve2003},
or to the (local) TDOS \cite{luther1974} of the wire, respectively.
The latter commonly displays power-law scaling dependent on the extent
of many-body interactions in the system~\cite{giamarchi,FisherGlazman1997}
-- quantified by the Luttinger parameter $K$ -- and is strongly
impacted by the boundaries, i.e. impedance of the environment~\cite{devoret2005}.

In this work, we study the impact of noisy capacitive metallic leads
on tunneling into an interacting quantum wire. The capacitance in
the leads imposes a finite response time in the wire, suppressing
its fast high-energy excitations. Surprisingly, with increasing many-body
interactions, the impact of the boundary noise on the wire is enhanced,
and its TDOS displays this interplay by entering a regime where it
is dominated by the classical impedance of the capacitive reservoirs:
at low energies, the finite length of the wire cuts off the expected
1D tunneling zero-bias anomaly~\cite{nazarov1997,FisherGlazman1997},
and a zero-bias tunneling peak appears instead as a function of the
environment capacitance; at high energies, the characteristic power-law
growth is replaced by a universal $\omega^{-3}$ decay~\cite{devoret2005}.
Interestingly, this wire--environment competition introduces a unique
spatial dependence to the TDOS, thus offering an external handle by
which to control the correlations in the wire, such that its Luttinger
parameter can be tunably detected.

\begin{figure}
\center
\includegraphics{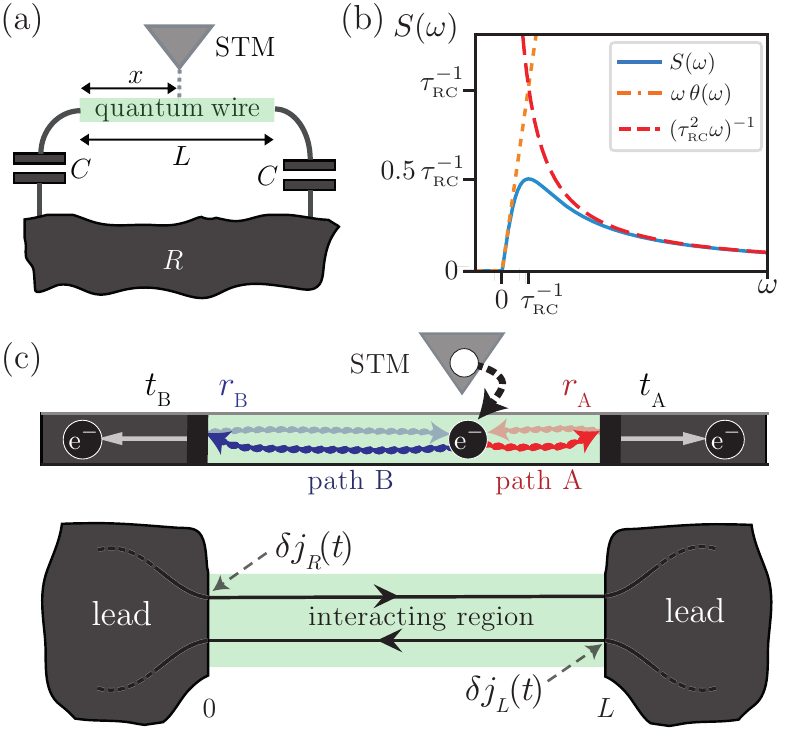}
\caption{ \label{main_fig:system_setting} (a) A 1D metallic quantum wire of
length $L$ is connected to metallic leads, depicted as an outer circuit
that is characterized by an ohmic resistance $R$ and the capacitance
$C$. The leads act as electron reservoirs with well-defined Fermi-Dirac
distributions. The tunneling density of states {[}Eq.~\eqref{main:tdos}{]}
at position $x$ along the wire is probed by a nearby scanning tunneling
microscope (STM). (b) The zero temperature power spectral-density
$S(\omega)$ of the RC-circuit's noise {[}Eq.~\eqref{main:RC_spectral_density_kernel}{]}
(blue solid line) and two asymptotic limits: (i) $\omega\,\tauRC\ll1$
(orange dot-dashed line) corresponding to the behavior of ideal ohmic
leads \cite{lifshitz_pitaevski1980} and (ii) $\omega\,\tauRC\gg1$
(red dashed line) where high-energy fluctuations are damped by the
circuit's capacitance. (c) (Top) An electron from the STM induces
1D plasmonic excitations, for which the finite wire acts as an effective
Fabry-Pérot interferometer with reflection (transmission) coefficients
$r_{{\rm A,B}}$ ($t_{{\rm A,B}}$). (Bottom) A schematic view of
the wire as left/right-propagating modes connected to two identical
leads, that impart current fluctuations $\delta j_{L/R}(t)$ onto
the wire {[}Eq.~\eqref{main:RC_spectral_density_kernel}{]}. }
\end{figure}

We consider a finite 1D wire coupled to metallic leads, depicted as
an outer circuit that is characterized by an ohmic resistance $R$
and the capacitance $C$, and probed by a nearby STM, see Fig~\ref{main_fig:system_setting}(a).
The STM signal measures the local TDOS at position $x$ along the
wire \cite{bruus_flensberg} 
\begin{align}
\nu(x,\omega)\!=i\!\!\int\dd t\,\e{i\omega t}\,\left(G^{>}(x,t)-G^{<}(x,t)\right),\label{main:tdos}
\end{align}
where $\omega$ is the electron's energy, and $G^{<}(x,t)$ and $G^{>}(x,t)$
are the lesser and greater Green's functions, respectively. We work
in natural units, where $\hbar,e=1$. In equilibrium, 
$G^{<}(x,t)=-G^{>}(x,-t)$ \cite{bruus_flensberg} and it suffices
to analyze $G^{<}(x,t)=i\,\expval{\psi^{\dagger}(x,t)\psi(x,0)}$,
where we wrote its definition using the electronic field-operator
$\psi(x,t)$, and the average is taken with respect to the equilibrium
ground state.

In $1$D, interacting electrons form a TLL with collective wave-like
plasmonic excitations~\cite{tomonaga1950,luttinger1963,haldane1981,giamarchi,vonDelft1998}.
An electron injected from the STM into the wire excites plasmonic
modes that propagate away such that the probability amplitude for
the excitation to tunnel back into the STM decreases faster than in
a non-interacting system. This decay manifests as a power-law in the
Green's function~\cite{vonDelft1998,ivan2012,gutman2010}

\begin{align}
\lim_{L\to\infty}G^{<}(x,t)=\frac{i\,\Lambda}{2\pi v_{F}}\frac{1}{(1+i\Lambda t)^{\alpha}}\,,\label{main:free_interacting_cases}
\end{align}
where $L$ is the length of the wire, $\Lambda$ is the bandwidth
of the electronic system, $v_{F}$ is the Fermi velocity, and $\alpha=\left(K+K^{-1}\right)/2\geq1$
is the interaction-dependent power-law exponent for the Luttinger
parameter $K$. For non-interacting systems $K=1$, and therefore
$\alpha=1$.

In a finite wire, the effects of many-body interactions compete with
the noise arising at the boundaries \cite{gutman2008,gutman2009,ivan2012,ivan2013}.
The latter is characterized, in our case, by a power spectral-density
\cite{machlup1954,assaf_oreg2012,shahmoon2018} 
\begin{align}
S(\omega) & \equiv\expval{\delta j_{L/R}(\omega)\delta j_{L/R}(-\omega)}=\frac{\omega\cdot(1-f_{FD}(\omega))}{1+\tauRC^{2}\omega^{2}}\,,\label{main:RC_spectral_density_kernel}
\end{align}
where $\tauRC=RC$ is the discharge time of the capacitor in the outer
circuit, and $f_{FD}(\omega)=(1+\exp[-\omega/k_{B}T])^{-1}$ is the
Fermi-Dirac distribution in the left and right leads -- assumed here
to be identical and uncorrelated. The main difference between \eqref{main:RC_spectral_density_kernel}
and the power spectral-density of ideal ohmic leads is that the RC-circuit
acts as an additional low-pass filter~\cite{ivan2012,lifshitz_pitaevski1980},
see Fig.~\ref{main_fig:system_setting}(b).

\begin{figure}
\center \includegraphics{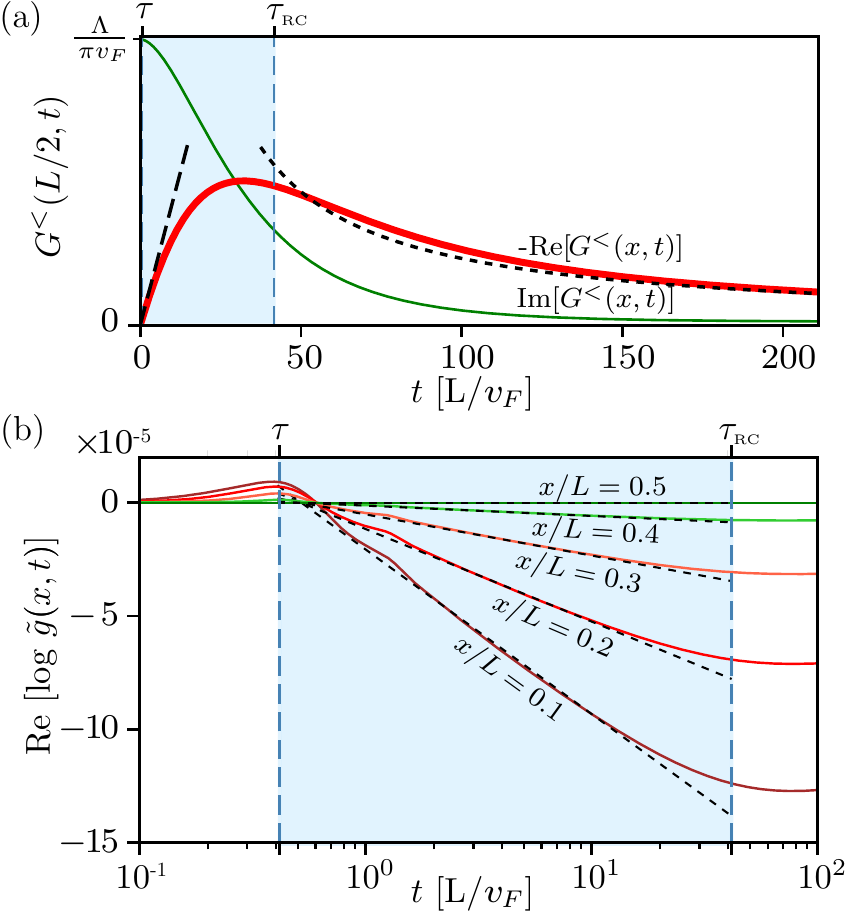} \caption{\label{main_fig:result} The Green's function of the wire in the strong-capacitance
limit. (a) The imaginary (thin green line) and real (thick red line)
part of a lesser Green's function $G^{<}(L/2,t)$ {[}Eq.~\eqref{main:correlator}{]}.
The dashed lines show the analytically obtained asymptotic limits
for long ($t\gg\tauRC$) and short ($t\ll\tauRC$) times. The shaded
region (light blue) marks the time interval $\tau<t<\tau_{RC}$ where
the interaction-induced correlations in the wire compete with the
RC noise. (b) The real part of $\log(\tilde{g}(x,t))$ {[}Eq.~\eqref{main:coherent_part}{]}
exhibiting the non-trivial power-law behavior of the Green's function
depending on the position of the STM tip (solid lines). Furthermore,
our analytical asymptotic result (dashed lines) {[}Eq.~\eqref{main:alpha}{]}
agrees with the numerical result (solid lines). In all plots, we use
an experimentally realizable interaction parameter $U/v_{F}=15$,
see, e.g., Refs. \cite{bockrath1999,yao1999}, and large capacitance,
$\tau_{RC}/\tau=100$. }
\end{figure}

We are interested in how the boundary noise~\eqref{main:RC_spectral_density_kernel}
and interaction-induced 1D plasmons manifest in the electronic correlations
in the wire, e.g., in $G^{<}(x,t)$. While the noise is characterized
by the discharge time $\tauRC$, we shall see below that the plasmonic
waves are characterized by their time-of-flight $\tau$ through the
finite wire, cf.~Eq.~\eqref{main:scattering_function}. We provide
here first a brief overview of our main results: the finite discharge
time of the leads imposes two distinct regimes, (i) \emph{strong-capacitance
regime} (see Fig.~\ref{main_fig:result}), where the time-of-flight
is much shorter than the discharge time, $\tau\ll\tauRC$, and (ii)
the more commonly studied complementary \emph{weak-capacitance regime}
with $\tau\gg\tauRC$. The latter shows a standard TLL behavior for
short times $t\leq\tau$, whereas for long times the finite wire acts
as a 0D Fabry-Pérot cavity for the plasmons and free-electron correlations
are reobtained (cf. Refs.~\cite{nazarov1997,devoret2005}). Case
(i) shows a richer behavior: at short times ($t\ll\tau,\tauRC$),
the boundary noise inhibits highly-excited plasmons and consequently
suppresses tunneling, whereas at long times ($t\gg\tau,\tauRC$),
both the interactions and noise correlations are averaged-out to yield
a similar 0D plasmonic Fabry-Pérot behavior. Interestingly, at intermediate
times ($\tau<t<\tauRC$), a competition between the TLL correlations
and the boundary response ensues, showing both Fabry-Pérot oscillations,
as well as non-trivial power-laws in the electronic correlations,
cf.~Eq.~\eqref{main:main_result} and see Fig.~\ref{main_fig:result}(a).
Furthermore, the power-laws show an unexpected dependence on the STM's
position \cite{Note4} that can be observed through {[}see Fig.~\ref{main_fig:result}(b){]}
\begin{align}
\tilde{g}(x,t)\equiv\frac{G^{<}(x,t)}{G^{<}(L/2,t)}\,.\label{main:coherent_part}
\end{align}

\begin{figure}
\flushleft \includegraphics{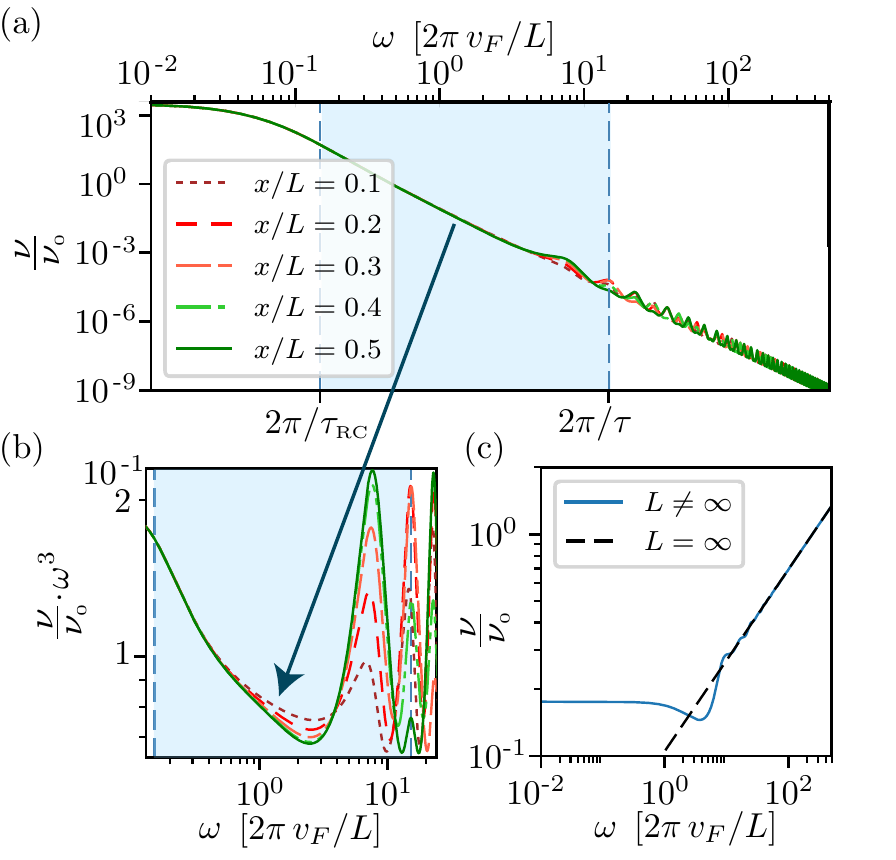} \caption{ \label{main_fig:TDOS} (a) The normalized TDOS $\nu/\nu_{0}$ in
the strong-capacitance regime ($\tauRC/\tau=100$) calculated for
five different STM positions in the wire. At high energies, $\omega\gg2\pi/\tauRC$,
the tunneling is suppressed and the TDOS exhibits a power-law decay.
Interaction-induced Fabry-Pérot oscillations with a period of $2\pi/\tau$
are present at high energies. For low energies, the TDOS is constant
that depends on the value of $\tauRC$. (b) A zoom-in on (a) where
the TDOS is rescaled by a factor $\omega^{3}$ such that the difference
between different measuring positions inside the wire can be seen
more clearly. (c) The TDOS of a finite (blue solid line) and infinite
(black dashed line) TLL when the capacitance in the leads is set to
zero. In a finite-length TLL, the zero-bias TDOS does not vanish but
saturates at a finite value \cite{nazarov1997}. Note that the normalization
of the TDOS is with respect to the value of non-interacting TDOS with
vanishing capacitance, $\nu_{0}$. The interaction strength used in
all plots is $U/v_{F}=15$. }
\end{figure}

In Fig.~\ref{main_fig:TDOS}(a), we plot the TDOS in the strong-capacitance
regime. The spatial dependence can be seen in the intermediate energy
regime, see Fig.~\ref{main_fig:TDOS}(b). For comparison, in Fig.~\ref{main_fig:TDOS}(c)
we plot the TDOS for both finite- and infinite-length interacting
wires. The relatively flat peak of the TDOS at low energies is a result
of the finite-length of the wire that suppresses the ZBA of an infinite
TLL {[}Fig.~\ref{main_fig:TDOS}(c){]}, and is in agreement with
the free-electron behavior of the Green's function at long times,
cf.~Fig.~\ref{main_fig:result}(a) and Refs.~\cite{nazarov1997,supmat}.
At high energies, interaction-induced Fabry-Pérot oscillations appear
but there is no interaction-dependent power-law growth as compared
with both the finite- and infinite-TLL, where the TDOS grows as $\nu(\omega)/\nu_{0}\propto\omega^{\alpha-1}$,
with $\alpha=(K+K^{-1})/2$ and $\nu_{0}=\nu(\omega,\tauRC=0,K=1)$
the TDOS into a non-interacting metal with zero capacitance. This
is a consequence of a linear, interaction-independent growth of the
Green's function at short times, see Fig.~\ref{main_fig:result}(a).
Hence, the noise of the capacitive leads suppresses the power-law
growth and causes the TDOS to drop as $\nu/\nu_{0}\propto\omega^{-3}$,
in similitude to high-impedance tunnel-junctions \cite{nazarov1992,devoret2005}.

To obtain our results, we closely follow the derivation used in Refs.~\cite{nazarov1997,supmat}.
We consider the Hamiltonian density of a single-channel wire \cite{vonDelft1998,giamarchi,gutman2010,ivan2012,nazarov1997}
\begin{align}
\mathcal{H}(x) & =-i\,v_{F}\,\left(\psi_{R}^{\dagger}(x)\,\px\psi_{R}(x)-\psi_{L}^{\dagger}(x)\,\px\psi_{L}(x)\right)\nonumber \\
 & \quad+\sum\limits _{\eta,\eta'=L,R}\int\dd y\,V_{\eta\eta'}(x-y)\rho_{\eta}(x)\rho_{\eta'}(y),
\end{align}
where the left- and right-moving electrons ($\eta=L,R$) are described
by field operators $\psi_{\eta}(x)$, and $V_{\eta\eta'}(x)$ is the
electronic interaction between (normal-ordered) density operators
$\rho_{\eta}(x)=\mathbf{:}\psi_{\eta}^{\dagger}(x)\psi_{\eta}(x)\mathbf{:}$.
The first term describes the kinetic contribution for a linearized
dispersion $E(\delta k)=v_{F}\,\delta k$ around the Fermi momentum
$k_{F}$, such that the electron field $\psi(x)\simeq\e{-ik_{F}x}\psi_{L}(x)+\e{ik_{F}x}\psi_{R}(x)$.
We further assume that the effective electron-electron interaction
is point-like, i.e. $V_{\eta\eta'}(x)=U\,\delta(x)$. Note that the
linearized dispersion is associated with a bandwidth $\Lambda$ serving
as a high-energy cut-off. Using bosonization~\cite{vonDelft1998},
we introduce new bosonic field operators $\phi_{\eta}(x)$ related
to the electron density by $\rho_{\eta}(x)=\partial_{x}\phi_{\eta}/2\pi$,
with commutation relations $\comm{\phi_{L/R}(x)}{\px\phi_{L/R}(y)}=\pm2\pi i\,\delta(x-y)$.
These fields are defined via $\psi_{\eta}(x)=:\hat{F}_{\eta}\left(\Lambda/[2v_{F}\pi]\right)^{1/2}\e{-i\phi_{\eta}(x)}$,
where the Klein factors $\hat{F}_{\eta}$ ensure fermionic anti-commutation
of $\psi_{\eta}$. In this language, the Hamiltonian takes a simple
quadratic form \cite{giamarchi,vonDelft1998,ivan2012} 
\begin{align}
\mathcal{H}(x) & \!=\!\left(\frac{v_{F}}{4\pi}+\frac{U}{8\pi^{2}}\right)\!\!\!\sum_{\eta={L,R}}\!\!(\px\phi_{\eta})^{2}\!+\!\frac{U}{4\pi^{2}}\,\px\phi_{L}\,\px\phi_{R}\,.\label{eq:Hamiltonian_Bosonized}
\end{align}
Substituting the bosonization identities into the lesser Green's function
of a finite wire, we obtain 
\begin{align}
G_{\eta}^{<}(x,t) & =\frac{i\,\Lambda}{2\pi v_{F}}\exp(\!-\frac{1}{2}\expval{(\phi_{\eta}(x,t)\!-\!\phi_{\eta}(x,0))^{2}})\,,\label{main:correlator}
\end{align}
where we have used the fact that the charge-fluctuations at the boundaries
are Gaussian distributed, and that $\expval{F_{\eta}^{\dagger}F_{\eta}}=1$.
Note that the overall Green's function is $G^{<}(x,t)=G_{L}^{<}(x,t)+G_{R}^{<}(x,t)$
\cite{bruus_flensberg}. Using the equations of motion for the fields
$\phi_{\eta}$ \cite{supmat}, we find (in similitude to Ref.~\cite{nazarov1997})
that $G_{\eta}^{<}(x,t)\equiv i\Lambda/(2\pi v_{F})\exp(-\mathcal{I}(x,t))$
with the integral 
\begin{align}
\mathcal{I}(x,t)=\int\limits _{-\Lambda}^{\Lambda}\frac{\dd\omega}{\omega^{2}}\,(1-\e{-i\omega t})\,\mathcal{F}(x,\omega)\,S(\omega)\,,\label{main:exponent_integral}
\end{align}
where $S(\omega)$ is as in Eq.~\eqref{main:RC_spectral_density_kernel}.
The structure-function 
\begin{align}
\mathcal{F}(x,\omega)\equiv\frac{1+\chi-2\chi\cos(\tau\omega)\cos(2\tau\omega\,(\frac{1}{2}-\frac{x}{L}))}{1-\chi\cos(2\tau\omega)}\label{main:scattering_function}
\end{align}
captures both interaction effects through the parameter $\chi^{-1}\equiv(1+8\pi\,v_{F}/U+8\pi^{2}\,(v_{F}/U)^{2})=[1-8K^{2}/(1+6K^{2}+K^{4})]^{-1}$,
and the finite-length of the wire through the time-of-flight of the
plasmonic excitations $\tau=(L/v_{F})\,(1+\pi^{-1}U/v_{F})^{-1/2}$.
This structure-function is equivalent to that of a plasmonic Fabry-Pérot
interferometer of length $L$. Indeed, the same expression is obtained
when describing a free-particle that is injected at a position $x$
and is reflected from the two boundaries with reflection and tunneling
coefficients $r_{\textrm{\tiny A,B}}\equiv r$, $t_{\textrm{\tiny A,B}}\equiv t$,
respectively, where $\chi=2\,r^{2}\,(1+r^{4})^{-1}$ {[}cf.~Fig.~\ref{main_fig:system_setting}(c)
and Refs.~\cite{nazarov1997,safi1995,gutman2009,Note2}{]}. This
implies that the plasmonic character of excitations in the wire (due
to interactions) causes reflections from the free-electron boundaries.

We can now (i) evaluate $G_{\eta}^{<}(x,t)$ numerically using Eqs.
\eqref{main:RC_spectral_density_kernel}~and~\eqref{main:correlator}-\eqref{main:scattering_function}
for different devices with varying $\tauRC/\tau$ and $U/v_{F}$ \cite{supmat},
as well as (ii) find analytical asymptotic results for the specific
time windows mentioned above. In the latter, we assume that the STM
is placed in proximity to the middle of the wire, such that $\left(1/2-x/L\right)\ll1$.

\paragraph*{Strong-capacitance regime~~$(\tau\ll\tauRC)$}

For short times, $t\ll\tau\ll\tauRC$, the real-part of the Green's
function is linear, while its imaginary-part reaches a finite value,
i.e., $G^{<}(x,t\rightarrow0)=\Lambda(\pi v_{F})^{-1}\,\left(i-\pi/2\cdot t/\tau_{\textrm{\tiny RC}}\right)$,
see Fig.~\ref{main_fig:result}(a). This behavior leads to the reduced
TDOS at high energies, see Eq.~\eqref{main:tdos} and Fig.~\ref{main_fig:TDOS}(a).
The large capacitance in the leads effectively acts as a low-pass
filter for the plasmonic modes, and inhibits the conversion of high-energy
STM electrons into plasmons.

At intermediate times, $\tau\ll t\ll\tauRC$, the main weight of the
integral $\mathcal{I}(x,t)$ {[}Eq.~\eqref{main:exponent_integral}{]}
lies at $\omega\gg\tauRC^{-1}$, where the spectral function is approximated
as $S(\omega)\approx1/\tauRC^{2}\cdot\omega^{-1}$. We expand the
cosine terms in Eq.~\eqref{main:scattering_function} in small $\tau/t\ll1$,
to obtain 
\begin{align}
G^{<}(x,t)\approx G^{<}\left(L/2,t\right)\cdot\frac{i\,\Lambda}{2\pi v_{F}}\frac{1}{(1+it\Lambda)^{\alpha(x)}}\,,\label{main:main_result}
\end{align}
with a spatially-dependent exponent 
\begin{align}
\alpha(x)=\left(\frac{1}{2}-\frac{x}{L}\right)^{2}\frac{(K^{2}-1)^{2}}{2K^{3}}\frac{\tau^{2}}{\tauRC^{2}}\,.\label{main:alpha}
\end{align}
The first factor in Eq.~\eqref{main:main_result} does not depend
on the position within the wire. Remarkably, however, the second factor
has the same power-law form as that of the Green's function of an
infinite TLL, see Eq.~\eqref{main:free_interacting_cases} -- with
the notable difference that the exponent has a spatial dependence.
This exponent can be extracted from $\tilde{g}(x,t)$ as defined in
Eq.~\eqref{main:coherent_part}, see Fig.~\ref{main_fig:result}~(b).

In the long time limit, $\tau\ll\tauRC\ll t$, the main weight of
the integral $\mathcal{I}(x,t)$ {[}Eq.~\eqref{main:exponent_integral}{]}
stems from small energies, $\omega\ll(\tauRC)^{-1}$, where the spectral
function is approximated as $S(\omega)\approx\omega\cdot(1-f_{FD}(\omega))$,
see Fig.~\ref{main_fig:system_setting}(b). Furthermore, for $\tau\ll t$,
the structure function is constant, i.e. $\mathcal{F}(x,\omega)\approx1+\mathcal{O}(\tau^{2}/t^{2})$.
Hence, the leading term in Eq.~\eqref{main:exponent_integral} becomes
$\mathcal{I}(x,t)=\gamma_{E}+\log(t/\tauRC)+i\pi/2$ with $\gamma_{E}$
the Euler constant, resulting in a free-electron response, $G^{<}(t)=-\Lambda(\pi v_{F})^{-1}\cdot\exp(\gamma_{E})\,\tauRC/t$
{[}cf. Eq.~\eqref{main:free_interacting_cases}{]}. The plasmons
created by the STM reflect back and forth multiple times between the
boundaries such that their interference 'washes out' the effects of
1D interactions, and a 0D plasmonic cavity forms \cite{nazarov1997,devoret2005}.

The interplay between noisy capacitive boundaries and many-body interactions
in a finite quantum wire can smoothly alter its temporal and spatial
correlations. Specifically, we find that the many-body interactions
drive the wire to display a TDOS with features that are dominated
by the classical fluctuations of its boundaries. Moreover, the emergent
TDOS is predicted to be spatially-dependent and can be measured using
a scanning tunneling microscope. Employing this emergent spatial-dependence
and control over the classical boundary noise, one can extract the
Luttinger parameter of a finite interacting wire with the ability
of preforming multiple measurements on a single sample. Our work opens
up interesting questions concerning the impact of an environment on
the TDOS into a wire, e.g., what would be the outcome of the competition
between the classical capacitive-noise studied here and strong out-of-equilibrium
noise~\cite{gutman2010}? A natural next step would be to investigate
the impact of treating quantum mechanical capacitive fluctuations.
Furthermore, another intriguing avenue would be to study similar correlations
in the context of modern synthetic atomic \cite{Cazalilla2011aa,Yang2017,Giamarchi2017}
and photonic \cite{Gullans2016} wires.

We would like to thank B. Rosenow, L. I. Glazman, I. V. Protopopov,
and I. V. Gornyi for fruitful discussions. We acknowledge financial
support from the Swiss National Science Foundation.

%


\newpage
\cleardoublepage
\setcounter{figure}{0}
\renewcommand{\figurename}{Supplementary Figure}

\onecolumngrid
\begin{center}
\textbf{\normalsize Supplemental Material for}\\
\vspace{3mm}
\textbf{\large  Tunneling into a finite Luttinger liquid coupled to noisy capacitive leads}
\vspace{4mm}

{ Antonio\ Štrkalj, Michael S. Ferguson, Tobias M. R. Wolf, Ivan Levkivskyi, and Oded Zilberberg}\\
\vspace{1mm}
\textit{\small Institute for Theoretical Physics, ETH Z\"urich, 8093 Z\"urich, Switzerland}

\vspace{5mm}
\end{center}
\setcounter{equation}{0}
\setcounter{section}{0}
\setcounter{figure}{0}
\setcounter{table}{0}
\setcounter{page}{1}
\makeatletter
\renewcommand{\bibnumfmt}[1]{[#1]}
\renewcommand{\citenumfont}[1]{#1}

\setcounter{enumi}{0}
\renewcommand{\theequation}{\Roman{enumi}.\arabic{equation}}

In the main text, we analyze a finite-length 1D wire subject to charge-fluctuations
at its boundaries {[}described by a power spectral-density $S(\omega)$
as in Eq. (3) in the main text{]}. The electronic modes in the wire are naturally described
by plasmonic modes according to the Hamiltonian density
\begin{equation*}
\mathcal{H}(x)=\left(\frac{v_{F}}{4\pi}+\frac{U}{8\pi^{2}}\right)\sum_{\eta=L,R}\left(\partial_{x}\phi_{\eta}\right)^{2}+\frac{U}{4\pi^{2}}\partial_{x}\phi_{L}\partial_{x}\phi_{R},
\end{equation*}
 where $\phi_{L,R}$ are bosonic fields, $v_{F}$ is the Fermi velocity
and $U$ is the (repulsive) interaction strength, see Eq. (6) in the
main text. The bosonic fields satisfy commutation relations $\comm{\phi_{\eta}(x)}{\partial_{x}\phi_{\eta}(y)}=\pm2\pi i\,\delta(x-y)$.

In Section I, we show how to obtain the eigenmodes of $\mathcal{H}(x)$
for given boundary conditions imposed by the current flucations at
the interface between the wire and the outer circuit. In Section
II and Section III we elaborate on the calculations for the weak and strong capacitance
regimes discussed in the main text. In Section IV, we show in more
detail how the real part of the wire's Green's function depends on
the length of the wire $L$ and the discharge time $\tau_{{\rm {\scriptscriptstyle RC}}}$ of the outer circuit's capacitor.
In Section V, we show how the TDOS behaves for different wire lengths $L$ and values of $\tauRC$.

\section{Equations of Motion}
\setcounter{enumi}{1} 
\setcounter{equation}{0}

The equations of motion for the modes in the wire can be obtained
from the Hamiltonian and the commutation relation for the left- and
the right-moving bosonic fields $\phi_{\eta}(x,t)$ \cite{nazarov}, i.e.,
\begin{align}
\pt\phix[L] & =i\comm{H}{\phix[L]}=\left(v_{F}+\frac{U}{2\pi}\right)\,\px\phix[L]+\frac{U}{2\pi}\px\phix[R],\label{EqOfMotion1}\\
\pt\phix[R] & =i\comm{H}{\phix[R]}=-\left(v_{F}+\frac{U}{2\pi}\right)\,\px\phix[R]-\frac{U}{2\pi}\px\phix[L].\label{EqOfMotion2}
\end{align}
The quadratic Hamiltonian density $\mathcal{H}(x)$ is diagonal in
the basis of the fields $\phix[\pm]=\frac{1}{\sqrt{2}}\left(\phix[L]\pm\phix[R]\right)$
, whose equations of motion directly follow from Eqs. \eqref{EqOfMotion1}
and \eqref{EqOfMotion2}, i.e., 
\begin{align}
\pt\phix[+] & =v_{F}\,\px\phix[-],\nn\label{SystemOfDiffEq}\\
\pt\phix[-] & =v_{F}\left(1+\frac{U}{\pi v_{F}}\right)\,\px\phix[+],
\end{align}
where the definition of the Luttinger parameter naturally emerge as
$K=\left(1+\frac{U}{\pi v_{F}}\right)^{-1/2}$.

The system of differential equations \eqref{SystemOfDiffEq} can be
solved in frequency space, i.e. $\pphi[\alpha]{x,t}=\frac{L}{2\pi v_{F}}\int\dd\omega\,\e{i\omega t}\phitt[\alpha]{x,\omega}$,
resulting in
\begin{align}
\px^{2}\phitt[\pm]{x,\omega}+\kappa(\omega)^{2}\phitt[\pm]{x,\omega} & =0\label{FourierEqOfMotion}
\end{align}
with a single parameter $\kappa(\omega)=K\cdot\omega/v_{F}$. The
solutions are plane waves of the form 
\begin{align}
\phitt[\pm]{x,\omega} & =c_{1,\pm}\e{i\,\kappa(\omega)\,x}+c_{2,\pm}\e{-i\,\kappa(\omega)\,x},\label{SolutionsForPlusMinus}
\end{align}
where the four coefficients $c_{1,\pm},c_{2,\pm}$ are still related
through Eqs.~\eqref{SystemOfDiffEq}, leading to the general solution
\begin{equation}
\left.\begin{array}{rl}
\phitt[+]{x,\omega} & =c_{1}\e{i\kappa x}+c_{2}\e{-i\kappa x}\\
\phitt[-]{x,\omega} & =c_{1}\frac{1}{K}\e{i\kappa x}-c_{2}\frac{1}{K}\e{-i\kappa x}
\end{array}\right\} \quad\Leftrightarrow\quad\left\{ \begin{array}{rl}
\phitt[L]{x,\omega} & =\frac{1}{\sqrt{2}}\left(c_{1}\gp\e{i\kappa x}+c_{2}\gm\e{-i\kappa x}\right)\\
\phitt[R]{x,\omega} & =\frac{1}{\sqrt{2}}\left(c_{1}\gm\e{i\kappa x}+c_{2}\gp\e{-i\kappa x}\right)
\end{array}\right.,\label{eq:SolutionsForPlusMinus2}
\end{equation}

where $\gamma_{\pm}=1\pm K^{-1}$ and $c_{1}$, $c_{2}$ are two independent
coefficients determined by boundary conditions. The latter are given
by the continuity equation for the fluctuating currents $\delta j_{L,R}(t)$
at the left and right reservoir {[}see Fig.~1(c) in the main text{]}
\begin{align}
\partial_{t}\phi_{L}(L,t)=2\pi\,\delta j_{L}(t),\quad\partial_{t}\phi_{R}(0,t)=2\pi\,\delta j_{R}(t).\label{BoundaryConditions}
\end{align}
Transforming Eq. \eqref{BoundaryConditions} to frequency space and
substituting the general solution Eq. \eqref{eq:SolutionsForPlusMinus2}
then results in
\begin{align}
 & \gp\e{i\kappa L}\,c_{1}+\gm\e{-i\kappa L}\,c_{2}=\frac{\sqrt{2}}{i\,\omega}\,\delta j_{L}(\omega)\nn\label{EquationsForCoefficients}\\
 & \gm\,c_{1}+\gp\,c_{2}=\frac{\sqrt{2}}{i\,\omega}\,\delta j_{R}(\omega)
\end{align}
where $\delta j_{L}(\omega),\delta j_{R}(\omega)$ are the Fourier
transforms of $\delta j_{L,R}(t)$. Solving for the coefficients $c_{1,2}$
and substituting them into Eq. \eqref{eq:SolutionsForPlusMinus2}
then yields
\begin{align}
\left(\begin{matrix}\phitt[L]{x,\omega}\\
\phitt[R]{x,\omega}
\end{matrix}\right) & =-i\frac{1}{\omega}\,\frac{1}{\gm^{2}\e{-i\kappa L}-\gp^{2}\e{i\kappa L}}\left(\begin{matrix}\gm^{2}\e{-i\kappa x}-\gp^{2}\e{i\kappa x} & -2i\gp\gm\sin(\kappa(L-x))\\
-2i\gp\gm\sin(\kappa x) & \gm^{2}\e{-i\kappa(L-x)}-\gp^{2}\e{i\kappa(L-x)}
\end{matrix}\right)\left(\begin{matrix}\delta j_{L}(\omega)\\
\delta j_{R}(\omega)
\end{matrix}\right).
\end{align}

\section{Weak capacitance regime~~$(\tau\gg\tauRC)$}
\setcounter{enumi}{2} 
\setcounter{equation}{0}

In the case where $\tau_{{\rm {\scriptscriptstyle RC}}}$ is the smallest
time scale, the spectral function of the boundary fluctuations [cf. Eq. (3) in the main text] is approximated as $S(\omega)=\omega\,(1-f_{FD}(\omega))$
{[}see Fig.~1(b) in a main text{]}. In this case the integral $\mathcal{I}(x,t)$
in Eq.~(9) of the main text can be evaluated in both asymptotic limits
$t\ll\tau$ and $t\gg\tau$.

In the short time limit ($t\ll\tau$), the fast-oscillating cosines
in Eq.~(10) of the main text can be averaged. The spatial dependance
near the middle of the wire ($x\approx L/2$) vanishes so that $\mathcal{F}(x,\omega)=\sqrt{(1+\chi)/(1-\chi)}\equiv\alpha$
and hence $\mathcal{I}(t)=\alpha\log(1+i\Lambda t)$. Combining the two approximations together, we obtain Eq.~(2) in the main text for the Green's function of the wire, i.e., 
\begin{align}
 & G^{<}(x\approx L/2,t\ll\tau)=\frac{i\,\Lambda}{2\pi v_{F}}\frac{1}{(1+i\Lambda t)^{\alpha}}.\label{main:G_finite_interacting}
\end{align}
This result can be interpreted as follows: an electron injected from
the STM forms plasmons, which for short times $t\ll\tau$ do not have
time to propagate to and reflect from the boundaries. Hence, there
is no Fabry-Pérot interference and we observe the Green's function behavior of (infinite) interacting 1D wires.

In the long time limit ($t\gg\tau$), the cosines in Eq.~(10) of
the main text can be expanded in small $\tau/t\ll1$ such that $\mathcal{F}(x,\omega)\approx1+\mathcal{O}(\tau^{2}/t^{2})$
and hence $\mathcal{I}(t)=\log(1+i\Lambda t)$. The Green's function
then takes the form 
\begin{align}
 & G^{<}(x\approx L/2,t\gg\tau)=\frac{i\,\Lambda}{2\pi v_{F}}\frac{1}{1+i\Lambda t}.\label{main:G_finite_free}
\end{align}
Comparing this result with Eq.~(3) in the main text, we recover the
result of free electrons ($\alpha=1$) similar to the long-time limit
of strong capacitance regime in the main text. This is not surprising,
since the long-time limit the emergent 0D Fabry-Perot cavity should always tend to the result of non-interacting electrons.

\section{Strong capacitance regime ~~$(\tau\ll\tauRC)$}
\setcounter{enumi}{3} 
\setcounter{equation}{0}

We can write the integral from Eq.~(9) of the main text as $\mathcal{I}(t)=\mathcal{I}_{\cos}(t)+\mathcal{I}_{\sin}(t)$
and evaluate it analytically in the limit of strong capacitance \citep{main:zwillinger,main_edvin_ivan2017}:
\begin{align}
\mathcal{I}_{\cos}(t)=\lim\limits _{\eta\rightarrow0}\int_{0}^{\infty}\dd\omega\,\frac{\omega}{\omega^{2}+\eta^{2}}\frac{1-\cos(\omega t)}{1+\tauRC^{2}\omega^{2}} & =\gamma_{E}+\log(\frac{t}{\tauRC})-\frac{\e{-t/\tauRC}}{2}\Ei[\frac{t}{\tauRC}]-\frac{\e{t/\tauRC}}{2}\Ei[-\frac{t}{\tauRC}],\\
\mathcal{I}_{\sin}(t)=-\lim\limits _{\eta\rightarrow0}\int_{0}^{\infty}\dd\omega\,\frac{\omega}{\omega^{2}+\eta^{2}}\frac{\sin(\omega t)}{1+\tauRC^{2}\omega^{2}} & =-\frac{\pi}{2}\left(1-\e{-t/\tauRC}\right),
\end{align}
where $\gamma_{E}=0.577...$ is an Euler's constant and $\Ei[x]=-\int_{-x}^{\infty}\dd y\,\e{-y}/y$
is the exponential integral for real non-zero values of $x$. The
asymptotic limits of the exponential integral $\Ei[t/\tauRC]$ are: 

\[
\Ei[\pm\frac{t}{\tauRC}]\simeq\begin{cases}
\gamma_{E}+\log(\frac{t}{\tauRC}) & \text{for \ensuremath{\quad}}t\ll\tauRC\\
\pm\frac{\e{\pm t/\tauRC}}{t/\tauRC} & \text{for \ensuremath{\quad}}t\gg\tauRC
\end{cases}
\]
This leads to the asymptotic result
\begin{align}
\mathcal{I}(t) & =\begin{cases}
i\frac{\pi}{2}\frac{t}{\tauRC} & \text{for }\quad t\ll\tauRC\\
\gamma_{E}+\log(\frac{t}{\tauRC})+i\frac{\pi}{2} & \text{for }\quad t\gg\tauRC
\end{cases}.
\end{align}

\section{Green's function as function of discharge time $\tauRC$ and length
$L$}
\setcounter{enumi}{4} 
\setcounter{equation}{0}

\begin{figure}[!h]
\center \includegraphics[scale=1.2]{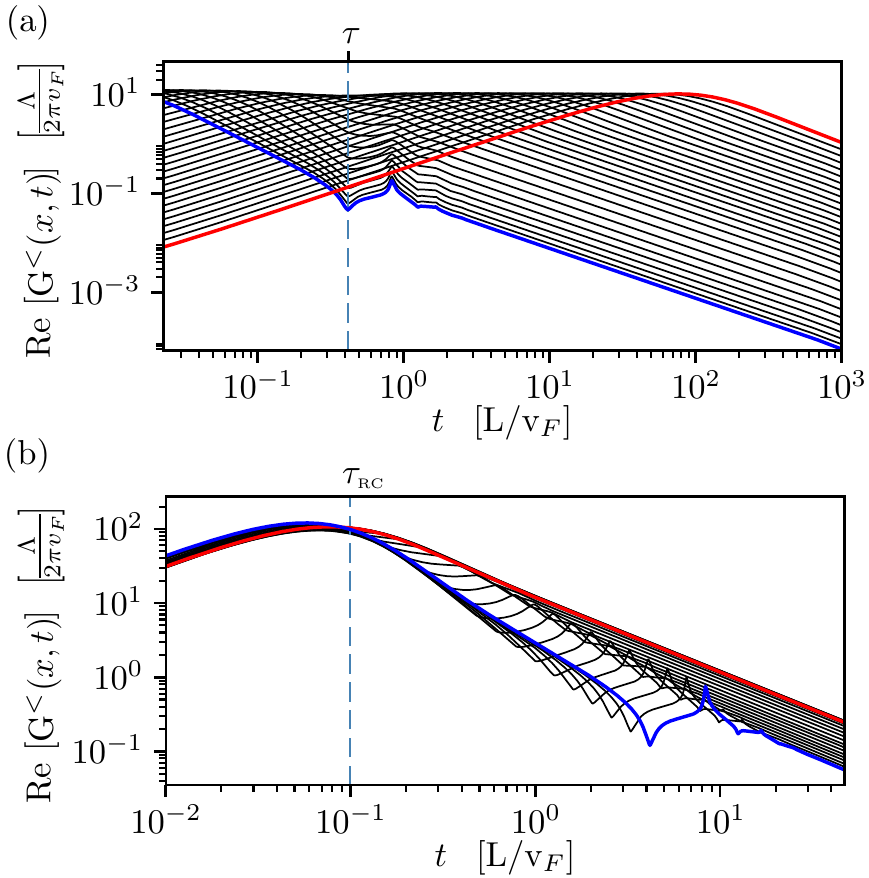} \caption{\label{fig:changing_RC} Dependence of the Green's function of a finite
wire embedded in a capacative circuit on (a) the discharge time $\tauRC$
and (b) the length $L$ of the wire. In both cases, the blue lines
mark the weak capacitance regime, $\tauRC/\tau\ll1$ and the red lines
mark the the strong capacitance regime $\tauRC/\tau\gg1$. The black
curves show how the Green's function interpolates between the two
regimes as the respective paramter is varied. In both plots,
interaction strength is fixed at $U/v_{F}=15$.}
\end{figure}

In Sup. Fig.~\ref{fig:changing_RC}(a), we show the behavior of the real
part of the Green's function $G^{<}(x,t)$ for different values of
the discharge time $\tauRC$ when the time of flight $\tau$ is fixed
(by fixing the length $L$ and interaction strength $U/v_{F}$). We
can see the smooth transition from the weak-capacitance regime to
the strong capacitance regime. The former is characterized by a $1/t^{\alpha}$
dependence at short times ($t\ll\tau$) due to the interactions and
a $1/t$ free electron behavior at long times ($t\gg\tau$). The latter shows interaction-independent linear dependence at short
times and a free electron behavior $1/t$ at long times. Furthermore,
in the weak capacitance limit, Fabry-Pérot oscillations with length-scale
$2\tau$ can be seen. In Sup. Fig.~\ref{fig:changing_RC}(b), we show
the same interpolation between weak- and strong-capacitance regime
but now keeping the value of the discharge time $\tauRC$ fixed while
instead changing the length $L$ of the wire (and consequently $\tau$).

\section{TDOS for different wire lengths $L$ and values of $\tauRC$}
\setcounter{enumi}{5} 
\setcounter{equation}{0}

\begin{figure}[!h]
	\center \includegraphics[scale=1.2]{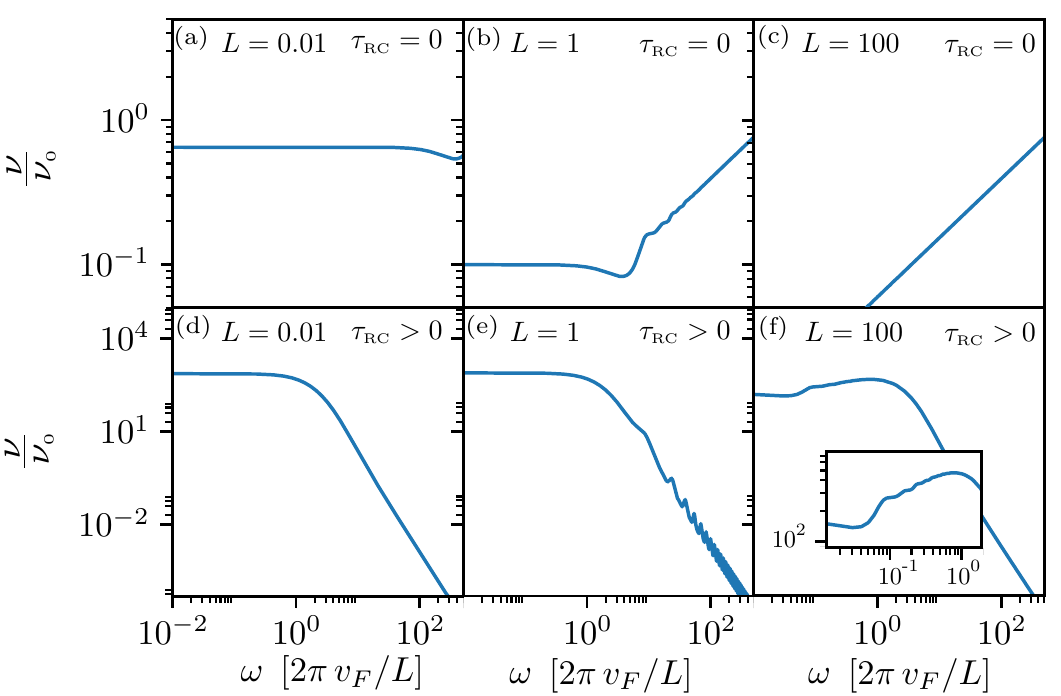} \caption{\label{fig:changing_L_many} 
	The normalized TDOS $\nu/\nu_{0}$ for
	different lengths $L$ {{[}in units of $v_F \, \tau${]}} of the
	wire for (a-c) in absence of an external capacative circuit ($\tauRC=0$)
	and (d-f) in presence of one ($\tauRC\protect\neq0$). The normalization
	is w.r.t. to the TDOS $\nu_{0}$ in the non-interacting case and in
	absence of the capacitance, i.e., $\tauRC=0$. (a) For a short wire,
	the constant TDOS indicates the free-electron behavior caused by multiple
	reflections of the plasmons against the boundaries. As $L$ is further
	decreased, the TDOS $\nu$ will approach $\nu_{0}$. (b) For intermediate
	lengths, the TDOS shows both free-electron behavior (constant TDOS)
	at low energies and TLL behavior (power-law growth) at high energies,
	see Fig. 3(c) in the main text. Furthermore, Fabry-Pérot oscillations
	with a period of $2 \pi / \tau$ appear. (c) For a long wire, the
	TDOS follows a power-law behavior characteristic for the infinite
	size TLL-s. In presence of a capacative outer circuit, the TDOS of
	(d) a short wire is completely determined by the fluctuations with an elevated zero-bias peak and universal $\omega^{-3}$ decay at high energies. 
	(e) At intermediate lengths, the result from the main text is recovered,
	see Fig. 3(a). Fabry-Pérot oscillations appear at higher energies,
	but the overall shape is the same as in the case of (d). (f) In case
	of a long wire, we see that the TDOS for low and high energies behaves
	as in (d), but in addition an intermediate regime appears, in which
	the power-law behavior of (b) is recovered {[}see also the inset
	of (f){]}. This intermediate region is characteristic for the tunneling  into a TLL. Throughout, we used a finite interaction
	strength $U/v_{F}=15$.}
\end{figure}

In the Sup. Fig.~\ref{fig:changing_L_many} we show the behavior of the tunneling density of states $\nu/ \nu_0$ for different lengths of the wire, both for vanishing discharge time, i.e., $\tauRC=0$ [Fig.~\ref{fig:changing_L_many}(a-c)] and for finite $\tauRC$ [Fig.~\ref{fig:changing_L_many}(d-f)].

\end{document}